\documentclass[reprint,showpacs,amsmath,amssymb,aps]{revtex4}
\usepackage{graphicx, subfigure}
\newcommand{\Array}[2]{\left(\begin{array}{#1}#2\end{array}\right)}

\begin{document}

\title{Decoding Fermion Mass Pattern from Mass Hierarchy}

\author{Ying Zhang$^{a,b}$\footnote{E-mail: hepzhy@mail.xjtu.edu.cn. }}
\address{$^a$School of Physics, Xi'an Jiaotong University, Xi'an, 710049, China}
\address{$^b$Institute of Theoretical Physics, Xi'an Jiaotong University, Xi'an, 710049, China}

\begin{abstract}
Building an organized Yukawa structure of quarks and leptons is an essential mission to understand fermion mass hierarchy and flavor mixing in particle physics.  Inspired by the similarity of CKM and PMNS mixings, a common mass pattern for up-type and down-type quarks, charged leptons, and Dirac neutrinos is realized in terms of hierarchal masses. 
An organized structure of Yukawa couplings can be expressed on a Yukawa basis.
The CKM mixing and PMNS mixing have the same mathematical forms due to $SO(2)$ family symmetry. By fitting the CKM/PMNS mixing experiment, the flat pattern becomes a successful flavor structure. The fundamental fermions can be explained by a new picture to show the relationship between gauge structure and flavor structure. 
\end{abstract}


\maketitle

\section{Motivation}
\label{sec.intro}

Compared with weak, strong, and electromagnetic gauge interactions that have been elegantly described by three simple couplings, Yukawa interactions are still ambiguous because of a great number of redundant couplings in the standard model (SM). 
After EWSB, Yukawa couplings generate fermions mass matrixes, which determine not only the physical masses of quarks but also charged leptons and quark/lepton flavor mixings in phenomenology \cite{FlavorRev}. However, the values of Yukawa couplings or even their structure have not been reconstructed from fermion masses and mixings. Finding an organized mass matrix becomes a fundamental question in particle physics, which is related to understanding the flavor nature and also seeking possible new physics.
Essentially, quark/lepton masses and $3\times 3$ unitary mixings all can be addressed from complex Yukawa couplings in the SM. Exploration of new physics beyond SM Yukawa couplings requires first demonstrating an organized structure of Yukawa couplings, followed by identification of the corrections that new physics provides outside the scope of SM Yukawa couplings.
Taking the quark sector as an example, all flavor parameters include 6 quark masses, 3 mixing angles, and 1 CP violating phase. Treating up-type and down-type quarks as independent requires 10 Yukawa coupling parameters. 
It brings the first challenge faced in flavor physics, to faithfully represent quark flavor structure in just 10 parameters. 

Another challenge in flavor structure is to understand the similarity between quark mixing and lepton mixing. Fermion mass matrix can be diagonalized by bi-unitary transformation $U_L^fM^f(U_R^f)^\dag={\rm diag}(m_1^f,m_2^f,m_3^f)$. 
Quark CKM mixing matrix defined as $U^u_L(U_L^d)^\dag$ and lepton PMNS mixing matrix defined as $U_L^e(U_L^\nu)^\dag$ have a similar mathematical form if considering the minimal extended SM with Dirac neutrinos. The similar mixing structures encourage us to build a common mass pattern for quarks and lepton. Facing numerical differences between two large mixing angles in the lepton PMNS matrix and small ones in the quark CKM matrix, a common mass pattern must provide a successful description of two kinds of mixing matrixes. 
In the lepton sectors, some concrete symmetries have been studied from the approximate structure of PMNS mixing \cite{DiscreteSym}. Because these researches focus on PMNS mixing special characteristics, it isn't easy to generalize these results to quarks.

As two sides of a flavor question, exploring the relation between fermion masses and mixings becomes an important clue to decoding flavor structure. Especially, the spectrum of up-type and down-type quarks, charged leptons, and extended normal-order Dirac neutrinos show characteristics of hierarchal mass $m_1^f\ll m_2^f \ll m_3^f$. Treating mass hierarchies as perturbations can reduce redundant parameters in complex mass matrices and reveal the underlying reason for flavor mixing.

In our previous paper \cite{ZhangJPG2023}, a flat flavor structure (with all elements of mass matrix unity) has been proposed as the common mass pattern of quarks and leptons, which arises from two assumptions. Now, we find these assumptions can completely be derived from the mass hierarchy condition. It concludes that the fermion mass pattern is determined by mass hierarchy essentially.   
The general mass pattern also provides a relation of two traditional patterns, i.e. flat pattern and seesaw pattern.

Almost all flavor data have been experimentally measured, except for the absolute masses of neutrinos and the lepton CP violation phase \cite{PDG2022}. 
The data from these flavor experiments have provided a benchmark for any models or structures related to flavors.

In the paper, we will derive a two-parameter mass pattern in the condition of the mass hierarchy in Sec. \ref{sec.pattern}. A general flavor mixing matrix is obtained. Flat pattern and seesaw pattern are pointed as the two special cases. In Sec. \ref{sec.breaking}, the flavor breaking in the flat and seesaw patterns is studied, respectively. 
We generalize all results of the quark sector to the lepton sector in Sec. \ref{sec.lepton}. 
Based on the mission of a common flavor structure in this paper, massive neutrinos are set as three normal-order Dirac fermions, which occupy all characteristics of the SM fermions, i.e. hierarchal masses and Dirac fermion. 
Flavor global fit to CKM and PMNS mixings are given in Sec. \ref{sec.fit} for flat pattern and seesaw pattern, respectively.
A summary is provided in Sec. \ref{sec.summary}, and a new fermion map is presented to clarify flavor and gauge structure.

\section{Mass Pattern in Hierarchy Limit}
\label{sec.pattern}
 
As an example, we will use quarks, but all results can be generalized to leptons directly in Sec.  \ref{sec.lepton}.

 Considering quark mass matrix $M^q$ (for $q=u,d$), it is diagonalized to mass eigenvalues by bi-unitary transformation 
\begin{eqnarray}
 U_L^qM^q(U_R^q)^\dag={\rm diag}(m_1^q,m_2^q,m_3^q) 
 \label{eq.UMU0001}
\end{eqnarray}
In phenomenology, all observable quantities are quark mass eigenvalues and CKM mixing. 
The latter is determined by only left-handed transformations $U_{CKM}=U_L^u(U_L^d)^\dag$.
For a random unitary $U'$, $M^qU'$ and $M^q$ have an identical phenomenology.
 It means that $M^q$ can not be speculated from measured massed and CKM mixing unless some principle is proposed to constrain right-handed $U_R^q$.

Let us choose $U_R^q=U_L^q$ and discuss mass patterns in the hierarchy limit. 
Defining mass ratio $h_{ij}^q=m_i^q/m_j^q$ ($i<j$), Eq. (\ref{eq.UMU0001}) becomes
\begin{eqnarray}
\frac{1}{\sum m_i^q}U_LM^q U_L^\dag 
= \frac{1}{h_{12}h_{23}+h_{23}+1}{\rm diag}(h_{12}h_{23},h_{23},1) 
\label{eq.UMUhie01}
\end{eqnarray}

Considering the mass hierarchy limit  $h_{ij}^q\ll 1$, RHS of Eq. (\ref{eq.UMUhie01}) reduces to ${\rm diag}(0,0,1)$. 
$M^q$ can be reconstructed in terms of $U_L^q$ as
\begin{eqnarray}
\frac{1}{{\sum m_i^q}}M^q
		=(U_L^q)^\dag{\rm diag}(0,0,1)U_L^q
\label{eq.U001U02}
\end{eqnarray}
We find that only three elements of the matrix, $(U_L^q)_{i3}$ for $i=1,2,3$, have non-vanishing contributions to $M^q$.

Parameterizing $(U_L^q)_{i3}$ by 3 real modulus and 3 real phases as 
		\begin{eqnarray*}
			(U_L^q)_{33}=l_0e^{i\eta_0},~~~
			(U_L^q)_{31}/(U_L^q)_{33}=l_1e^{i\eta_1},~~~
			(U_L^q)_{32}/(U_L^q)_{33}=l_2e^{i\eta_2}
		\end{eqnarray*}
we have $l_0^2(1+l_1^2+l_2^2)=1$  in terms of  the unitary of $U_L$.
All complex phases in Eq. (\ref{eq.U001U02}) can be factorized into diagonal $K_L^q$
	\begin{eqnarray}
		\frac{1}{\sum m_i^q}M^q
		=(K_L^q)^\dag
		M_N^q
		K_L^q
	\label{eq.KMK2}
	\end{eqnarray}
with $K_L^q\equiv {\rm diag}\Big(e^{i\eta_1},e^{i\eta_2},1\Big)$. 
Here normalized matrix $M_N^q$ determines the mass pattern
\begin{eqnarray}
M_N^q=\frac{1}{l_1^2+l_2^2+1}\Array{ccc}{l_1^2 & l_1l_2 & l_1 \\
			l_1l_2 & l_2^2 & l_2 \\
			l_1 & l_2 & 1
		} 
\label{eq.PatternM}
\end{eqnarray}
which meets $tr[M_N]=1$. (In the following text, we will ignore the superscript $^q$ when there is no possibility of confusion.)

 As we know, Yukawa interactions in the SM are not determined by the gauge principle. When describing Yukawa interaction in terms of gauge eigenstate, there may exist a mismatch of eigenstates. 
 It is possible to define Yukawa interaction on a new basis to exhibit an organized structure of couplings \cite{ZhangJPG2023}. 
It has been realized by the factorized form of Eq. (\ref{eq.KMK2}).
Defining a new Yukawa basis in terms of $K_L$
	\begin{eqnarray}
		q_i^{(Y)}=(K_L)_{ij} q_j 
	\nonumber
	\end{eqnarray}
with family index $i,j$, the Yukawa term after EWSB can be expressed into a family universal from
 	\begin{eqnarray}
		-\mathcal{L}_m=\left(\sum_i m_i^q\right)\bar{q}_L^{(Y)} M_Nq_R^{(Y)} 
	\nonumber
	\end{eqnarray}

In $U_{CKM}$, non-vanishing $(K_L^u)^\dag K_L^d$ provides all complex phases required by the CP violating phase. It means that the origin of CP-violating can be explained in terms of the transformation of gauge basis and Yukawa basis.

The $M_N$ has a $SO(2)$ family symmetry (also called as horizontal symmetry) in the space of families.
Defining Hermitian operator 
\begin{eqnarray}
	T=\frac{1}{\sqrt{l_1^2+l_2^2+1}}\Big(l_1 T_1+l_2 T_2+ T_3\Big)
	\nonumber
\end{eqnarray}
with three $SO(3)$ generators $T_i$ for $i=1,2,3$,
we have 
\begin{eqnarray}
 [T, M_N]=0 
	\nonumber
\end{eqnarray}
It means that $M_N$ is invariant under a $SO(2)$ rotation $R_N(\theta)=e^{i\theta T}$ along the direction  of $N=(l_1,l_2,1)$.

The diagonalization of $M_N$ can generally be realized by the product of two orthogonal transformations $SR_N(\theta)$
\begin{eqnarray}
	&&\Big[ SR_N(\theta)\Big]M_N \Big[SR_N(\theta)\Big]^T=SM_NS^T={\rm diag}(0,0,1)
	\label{eq.SRMRS}
\end{eqnarray}
with
\begin{eqnarray}
	S&\equiv & \Array{ccc}{\frac{1}{\sqrt{1+l_1^2}} & 0 & -\frac{l_1}{\sqrt{1+l_1^2}} \\
		-\frac{l_1l_2}{\sqrt{(1+l_1^2)(1+l_1^2+l_2^2)}} & \frac{\sqrt{1+l_1^2}}{\sqrt{1+l_1^2+l_2^2}} & -\frac{l_2}{\sqrt{(1+l_1^2)(1+l_1^2+l_2^2)}} \\
		\frac{l_1}{\sqrt{1+l_1^2+l_2^2}} & \frac{l_2}{\sqrt{1+l_1^2+l_2^2}} & \frac{1}{\sqrt{1+l_1^2+l_2^2}} }
	\label{eq.generalSexpression}
\end{eqnarray}
The above rotation $R_N(\theta)$ is just $R_3(\theta)$ on the basis of mass eigenstates because of 
\begin{eqnarray}
SR_N(\theta)S^T=R_3(\theta)=\Array{ccc}{c_\theta & s_\theta & 0 \\ -s_\theta & c_\theta & 0 \\ 0 & 0 & 1}
\nonumber
\end{eqnarray}

Thus, Eq. (\ref{eq.SRMRS}) can alternatively be expressed as
\begin{eqnarray}
\Big[R_3(\theta)S\Big]M_N\Big[R_3(\theta)S\Big]^T={\rm diag}(0,0,1)
\label{eq.RMRdiag0}
\end{eqnarray}

Using Eqs. (\ref{eq.KMK2}) and (\ref{eq.RMRdiag0}), the unitary $U_L^q$ that transforms left-handed quark from gauge eigenstates to mass eigenstates is 
\begin{eqnarray}
	U_L^q=R_3(\theta^q)S\Big[K_L^q\Big]^\dag
\label{eq.Ulq01}
\end{eqnarray}
So, the CKM mixing matrix is expressed as
\begin{eqnarray}
	U_{CKM}=U_L^u(U_L^d)^\dag=R_3(\theta^u)S ~{\rm diag}(e^{i\lambda_1},e^{i\lambda_2},1)S^TR_3^T(\theta^d)
\label{eq.Uckm0}
\end{eqnarray}
with Yukawa phase $\lambda_i=-\eta_i^u+\eta_i^d$.
Obviously, obserable Yukawa phases are only two $\lambda_1$ and $\lambda_2$ rather than all four phases $\eta_{1,2}^{u,d}$ in $K_L^{u,d}$.
We also notice that
$SO(2)$ rotation angles $\theta^{u},\theta^d$ play a non-trivial role in $U_{CKM}$ \cite{XuEPL2023}. 
Now,
$U_{CKM}$  includes four free parameters to address the same number of mixing parameters \cite{ChauPRL1984}.
It has responded to the first challenge in the hierarchy limit: non-redundant parameters in the mixing matrix. 

In flavor models, two types of mass patterns are frequently discussed:
\begin{itemize}
\item[1.]
$l_1=l_2=1$: flat pattern 
\begin{eqnarray}
M_N^{flat}=\frac{1}{3}\Array{ccc}{1&1&1\\ 1& 1 & 1\\ 1& 1& 1}
\label{eq.MNFlat}
\end{eqnarray}
A similar pattern is known as the democratic matrix \cite{Democratic}. In fact, the flat pattern is different from the democratic type in two major aspects. Flat patterns are constructed based on a Yukawa basis. All complex phases are separated into $K_L^{u,d}$, so CP violation should not be considered when breaking flavors. However, complex corrections must be introduced into the democratic type.  Another difference is that flat pattern treats up-type and down-type quarks in the same fashion.
\item[2.]
$l_1=l_2=0$: seesaw pattern \cite{seesawPapers}
\begin{eqnarray}
M_N^{seesaw}=\Array{ccc}{0 & 0 & 0 \\ 0 & 0 & 0 \\ 0 & 0 &1}
\label{eq.MNSeesaw}
\end{eqnarray}
\end{itemize}

$M_N^{flat}$ explains hierarchal mass from a non-distinguished structure, while $M_N^{seesaw}$ presupposes a hierarchyal structure.

Notice that we have set $U_R=U_R$ based on the non-observable role of $U_R$ in phenomenology. Otherwise, the two-parameter mass pattern would be applicable for $M_N(M_N)^\dag$, but not for $M_N$. In fact, for the flat pattern and seesaw pattern, $M_N(M_N)^\dag$ and $M_N$ are the same matrix, i.e. 
$ (M_N)(M_N)^\dag=(M_N)
$ for $M_N=M_N^{flat}$ and $M_N^{seesaw}$.

\section{$M_N$ breaking}
\label{sec.breaking}
Before verifying these two patterns, flavor breaking must be considered. One reason is that mass hierarchy is also flavor phenomenology besides quark/lepton mixings. However, more importantly, it is related to how to understand the small mixing angles in the CKM and PMNS matrixes. In some flavor models, the small mixing angles are explained from hierarchy corrections \cite{FritzschHierarchy}.  This scheme has to face a choice to describe mass matrixes of quarks and leptons in different patterns to generate large PMNS mixing angles. 
To realize a common mass structure for quarks and leptons in mathematical form, both large PMNS mixing angles and small CKM ones should be addressed from a mechanism independent of mass hierarchies. 
So, the hierarchy correction to flavor mixing must be calculated before the global fit of flavor mixing.

A clue to breaking flavor can be found in the two-parameter mass pattern in Eq. (\ref{eq.PatternM}).
Any flavor breaking will make the two-parameter relation lose effectiveness.
For non-vanishing $l_1,l_2$, we can define $l_1$ and $l_2$ in terms of diagonal elements $M_{N,11}$ and $M_{N,22}$ as
\begin{eqnarray}
l_1^2=\frac{M_{N,11}}{M_{N,33}}
,~~~
l_2^2=\frac{M_{N,22}}{M_{N,33}}
	\nonumber
\end{eqnarray}
which keeps $tr[M_n]=1$.
Thus, the invalidation of the two-parameter relation manifests 3 symmetric non-diagonal corrections
$\delta_{12},\delta_{23}$ and $\delta_{13}$ as
\begin{eqnarray}
		M_{N}=\frac{1}{l_1^2+l_2^2+1}\Array{ccc}{l_1^2 & l_1l_2+\delta_{12} & l_1+\delta_{13} \\
			l_1l_2+\delta_{12} & l_2^2 & l_2+\delta_{23} \\
			l_1+\delta_{13} & l_2+\delta_{23} & 1
		}
\label{eq.MNdelta00}
\end{eqnarray}
Because only two free parameters are required to yield two hierarchies, $h_{12}$ and $h_{23}$, there must exist a family symmetry on three $\delta_{ij}$ corrections.

\subsection{Flat Pattern Breaking}

Now, we consider a broken flat pattern 
\begin{eqnarray}
M_N^{flat}=\frac{1}{3}\Array{ccc}{1 & 1+\delta_{12} & 1+\delta_{13} \\
1+\delta_{12} & 1 & 1+\delta_{23} \\
1+\delta_{13} & 1+\delta_{23} & 1}
\label{eq.brokenFlat0a}
\end{eqnarray}

To generate the eigenvalues $(0,h_{23},1-h_{23})$ in $\mathcal{O}(h^1)$,  
$\delta_{ij}$  satisfy the following relation \cite{XuEPL2023}
	\begin{eqnarray}
		\delta_{12}(\theta)&=&\Big(-\frac{3\cos(2\theta)}{4}-\frac{9\sin(2\theta)}{4\sqrt{3}}-\frac{3}{2}\Big)h_{23}+\mathcal{O}(h^2)
		\label{Eq.deltaTransdelta12}
		\\
		\delta_{23}(\theta)&=&\Big(-\frac{3\cos(2\theta)}{4}+\frac{9\sin(2\theta)}{4\sqrt{3}}-\frac{3}{2}\Big)h_{23}+\mathcal{O}(h^2)
		\label{Eq.deltaTransdelta23}
		\\
		\delta_{13}(\theta)&=&\Big(\frac{3\cos(2\theta)}{2}-\frac{3}{2}\Big)h_{23}+\mathcal{O}(h^2)
		\label{Eq.deltaTransdelta13}
	\end{eqnarray}
These relations are just the manifest of the $SO(2)$ family symmetry in Eq. (\ref{eq.RMRdiag0}) with 1-order hierarchy correction.

Taking initialed $\theta=0$, we have
\begin{eqnarray}
	(\delta_{12},\delta_{23},\delta_{13})=-\frac{9}{4}h_{23}\Big(1,1,0\Big)
	\label{eq.FlatPinitialValue}
\end{eqnarray}
Labeling $M_N^{flat}(\theta)$ generated by $\delta_{ij}(\theta)$ in Eqs. (\ref{Eq.deltaTransdelta12})-(\ref{Eq.deltaTransdelta13}), $M_N^{flat}(\theta)$ can be expressed by a $SO(2)$ rotation $R_\delta(\theta)$ along the direction $(1,1-\frac{9}{4}h_{23},1)$
	\begin{eqnarray}		
		M_N^{flat}(\theta)=R_\delta(\theta)M_N^{flat}(0) [R_\delta(\theta)]^T
	\label{eq.MthetaM0}
	\end{eqnarray}
$M_N^{flat}(0)$ can be diagonalized by $S_{\delta}$
	\begin{eqnarray}		
S_{\delta}M_N^{flat}(0)S_{\delta}^T={\rm diag}\Big(0,h_{23},1-h_{23}\Big)
	\label{eq.SM0S32}
	\end{eqnarray}
In $\mathcal{O}(h_{ij}^1)$, $S_\delta$ can be written into $S_0$ and its hierarchy correction	
	\begin{eqnarray}
		S_{\delta}=S_0+\frac{1}{4\sqrt{3}}h_{23}\Array{ccc}{0&0&0\\ \sqrt{2}&\sqrt{2} &\sqrt{2} \\ 1 & -2 & 1}+\mathcal{O}(h^2)
	\nonumber
	\end{eqnarray}
Here, $S_0$ transforms unbroken $M_N^{flat}$ to ${\rm diag}(0,0,1)$
	\begin{eqnarray}
		\frac{1}{3}S_0 \Array{ccc}{1 & 1& 1\\1 & 1& 1\\1 & 1& 1}S_0^T={\rm diag}(0,0,1)
		,~~~
		S_0=\Array{ccc}{\frac{1}{\sqrt{2}} & 0 & -\frac{1}{\sqrt{2}} \\ 
			-\frac{1}{\sqrt{6}} & \sqrt{\frac{2}{3}} & -\frac{1}{\sqrt{6}} \\
			-\frac{1}{\sqrt{3}} & \frac{1}{\sqrt{3}} & \frac{1}{\sqrt{3}}}
		\label{eq.S0expression}
	\end{eqnarray}

Using Eqs. (\ref{eq.MthetaM0}) and (\ref{eq.SM0S32}), the diagonalization of $M_N(\theta)$ can be realized by the transformation $S_\delta [R_\delta(\theta)]^T$
as 
\begin{eqnarray}
		\Big[S_\delta [R_\delta(\theta)]^T\Big]M_N^{flat}(\theta)\Big[R_\delta(\theta)(S_\delta)^T\Big]
		={\rm diag}(0,h_{23},1-h_{23})+\mathcal{O}(h^2)
	\label{eq.SO2symFlatP}
\end{eqnarray}

Finally, the quark mixing matrix including 1-order hierarchy correction is expressed by
	\begin{eqnarray}
		U_{CKM}^{flat}=\Big[S^u_\delta [R_{\delta}(\theta^u)]^T\Big]\Big[{\rm diag}( e^{i\lambda_1},e^{i\lambda_2},1)\Big]
		\Big[R_{\delta}(\theta^d)(S^d_\delta)^T\Big]
	\label{eq.ClostToFlatCKM01}
	\end{eqnarray}

From Eq. (\ref{eq.Uckm0}) in hierarchy limit to Eq. (\ref{eq.ClostToFlatCKM01}) with hierarchy correction, $U_{CKM}^{flat}$ is always dominated by two $SO(2)$ rotation angles and two Yukawa phases.  Mass hierarchies $h_{23}^{u,d}$ only contribute small corrections \cite{XuEPL2023}.

\subsection{Seesaw Pattern Breaking}
In $\mathcal{O}(h_{ij}^1)$, the seesaw pattern $M_N^{seesaw}$ has eigenvalues $(0,h_{23},1-h_{23})$. 
By applying the transformation $S_0$, the diagonal seesaw pattern can be converted to the flat pattern, which allows the results obtained for the flat pattern to be used for breaking the seesaw pattern.

Defining
\begin{eqnarray}
 M_N^{seesaw}(0)
&=&
S_0M_N^{flat}(0)S_0^T
\nonumber\\
&=&{\rm diag}(0,0,1)
		+\Array{ccc}{0 & 0 & 0 \\
		0 & 1 & -\frac{1}{2\sqrt{2}} \\
		0 & -\frac{1}{2\sqrt{2}} & -1}h_{23}
	\label{eq.1orderSeesaw}
	\end{eqnarray}
it gives a broken seesaw pattern that corresponds to $\theta=0$ in the flat pattern.

With the help of $SO(2)$ symmetry in Eq. (\ref{eq.MthetaM0}) for the flat pattern, a rotated seesaw pattern, labeled by $M_N^{seesaw}(\theta)$, exhibits the similar $SO(2)$ symmetry
	\begin{eqnarray}
		M_N^{seesaw}(\theta)&=&S_0M_N^{flat}(\theta)S_0^T
		\nonumber\\
		&=&R'_{\delta}(\theta)M_N^{seesaw}(0)[R'_{\delta}(\theta)]^T
		\label{eq.DefM1deltatheta}
	\end{eqnarray}
with 
	\begin{eqnarray}		
	R'_{\delta}(\theta)&=&S_0R_{\delta}(\theta)S_0^T
	\nonumber
	\end{eqnarray}
Obviouly, $M_{N}^{seesaw}(\theta)$ has the invariant eigenvalues $(0,h_{23},1-h_{23})$ in $\mathcal{O}(h_{ij}^1)$.
By this method, the $M_N^{seesaw}$ provides an 
equivalent breaking form with $M_N^{flat}$, which is a reappearance of Eq. (\ref{eq.brokenFlat0a}) in a new basis.
Notice that besides real parameterization, the seesaw pattern here
is also different from the traditional one in the treatment of up-type and down-type quarks. Two kinds of quarks are considered to have the same mass pattern in our paper.

Using Eq. (\ref{eq.SO2symFlatP}), we get a general orthogonal transformation to diagonalize $M_N^{seesaw}(\theta)$ as 
\begin{eqnarray}
		\Big[S_{\delta} [R_{\delta}(\theta)]^TS_0^T\Big] 
			M_N^{seesaw}(\theta)
			\Big[S_0R_{\delta}(\theta)(S_{\delta})^T\Big]
		=
		{\rm diag}(0,h_{23},1-h_{23})+\mathcal{O}(h^2)
		\label{eq.m1diagonalization}
	\end{eqnarray}
Now, the CKM mixing matrix in 1-order hierarchy correction can be expressed into
	\begin{eqnarray}
		U_{CKM}^{seesaw}=\Big[S^u_{\delta} [R^u_{\delta}(\theta^u)]^TS_0^T\Big]\Big[{\rm diag}(1, e^{i\lambda_1},e^{i\lambda_2})\Big]\Big[S_0R^d_{\delta}(\theta^d)(S^d_{\delta})^T\Big]
	\label{eq.ClostToseesawCKMH1}
	\end{eqnarray}

\section{lepton flavor structure}
\label{sec.lepton}

Neutrino masses must be introduced to the SM for neutrino oscillation. As electrically neutral fermions, neutrinos have two distinct natures, namely Dirac and Majorana.
Compared with the Majorana type, the Dirac neutrinos share more common properties with other SM fermions, which makes it easy to build a common flavor structure for all quarks and leptons. 
In our paper,  normal-order Dirac neutrinos are assumed to generalize quark flavor structure to leptons.
Using measured squared mass differences $\Delta m_{21}^2$ and $\Delta m_{32}^2$, neutrino masses are dependent on an initial value, i.e. the first family $m_1^\nu$.

Following the similar way to reconstruct mass matrixes of charged leptons and neutrinos in the hierarchy limit as Eqs. (\ref{eq.U001U02}) and (\ref{eq.KMK2}) of quarks, 
all complex phases in the mass matrix can be factorized into diagonalized masses $K_L^{l}$ (for $l=e,\nu$). Normalized lepton mass matrix $M_N^{l}$ can be expressed by a two-parameter pattern as shown in Eq. (\ref{eq.PatternM}). 	The $SO(2)$ flavor symmetry is also valid in the lepton sector.
So, the PMNS mixing matrix has the same mathematical structure as CKM mixing in Eq. (\ref{eq.Uckm0}) only by replacing superscript index $u\rightarrow e, d\rightarrow \nu$.

Chosen the value of $l_1^l=l_2^l=1$, the flat patterns of leptons are gotten.  Lepton flavor breaking can be considered in the same way as Eq. (\ref{eq.brokenFlat0a}). The PMNS mixing matrix with 1-order hierarchy correction becomes
\begin{eqnarray}
		U_{PMNS}^{flat}=\Big[S^e_\delta [R_{\delta}(\theta^e)]^T\Big]\Big[{\rm diag}( e^{i\lambda_1},e^{i\lambda_2},1)\Big]
		\Big[R_{\delta}(\theta^\nu)(S^\nu_\delta)^T\Big]
	\label{eq.ClostToFlatPMNSflatH1}
	\end{eqnarray}
The seesaw pattern of lepton can be gotten by taking $l_1^l=l_2^l=0$. Using the same scheme, the PMNS mixing matrix of the seesaw pattern with 1-order hierarchy correction becomes
	\begin{eqnarray}
		U_{PMNS}^{seesaw}=\Big[S^e_{\delta} [R_{\delta}(\theta^e)]^TS_0^T\Big]\Big[{\rm diag}(1, e^{i\lambda_1},e^{i\lambda_2})\Big]\Big[S_0R_{\delta}(\theta^\nu)(S^\nu_{\delta})^T\Big]
		\label{eq.ClostToseesawPMNSH1}
	\end{eqnarray}
Now, our second motivation has also been realized: a common mixing structure in mathematical form for quarks and leptons. All results are based on the hierarchy structure of the masses.
The similar forms of mixing matrix in the quark and lepton sectors show that CP violations in CKM and PMNS matrixes have the same origin, namely Yukawa phases.
The lepton hierarchy only plays a small correction. It means that the two large PMNS mixing angles must come from $SO(2)$ rotation angles and Yuakwa phases rather than the mass hierarchy of neutrinos or charged leptons \cite{XuEPL2023}. 
It is a coming test to yield two large mixing angles of PMNS mixings from  Eqs. (\ref{eq.ClostToFlatPMNSflatH1}) and (\ref{eq.ClostToseesawPMNSH1}).

\section{Flavor Mixing Global Fit}
\label{sec.fit}

Currently, CKM/PMNS mixing data have been measured from nuclear weak beta decays and neutrino oscillations \cite{PDG2022}. These precise mixing angles become a checkpoint to possible flavor structures proposed from different motivations.
In Eqs.  (\ref{eq.ClostToFlatCKM01}), (\ref{eq.ClostToseesawCKMH1}), (\ref{eq.ClostToFlatPMNSflatH1}) and (\ref{eq.ClostToseesawPMNSH1}), flavor mixing matrixes of quarks and leptons for the flat pattern and seesaw pattern with hierarchy corrections have been derived. What is left to do is to check the validity of flavor mixings by fitting $U_{CKM}$ and $U_{PMNS}$ to experiment data.

\begin{figure}[htbp]
	\centering  
	\subfigure[flat pattern to CKM in $\theta^u$-$\theta^d$]
		{\includegraphics[scale=0.25]{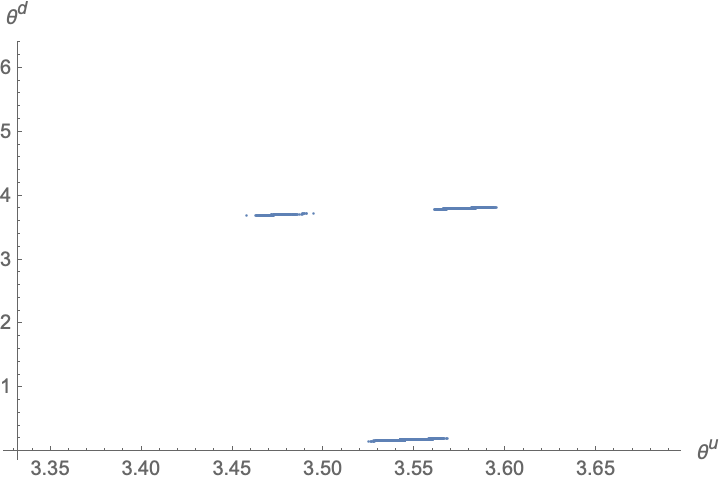}} ~~~~
	\subfigure[flat pattern to CKM in $\lambda_1$-$\lambda_2$]
		{\includegraphics[scale=0.25]{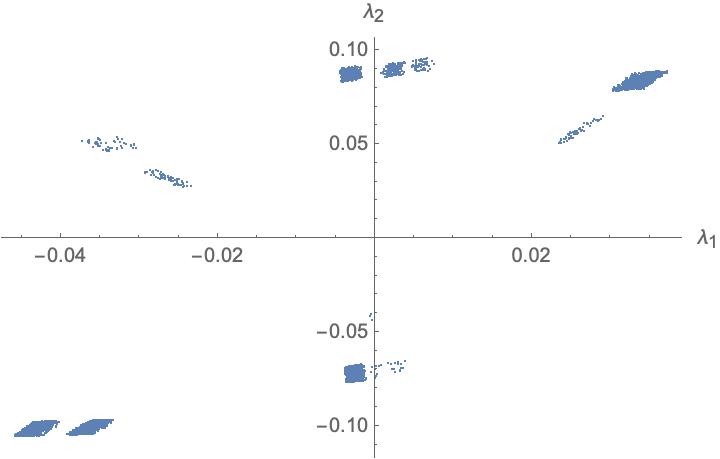}}
	\\
	\subfigure[flat pattern to PMNS in $\theta^e$-$\theta^\nu$]
		{\includegraphics[scale=0.25]{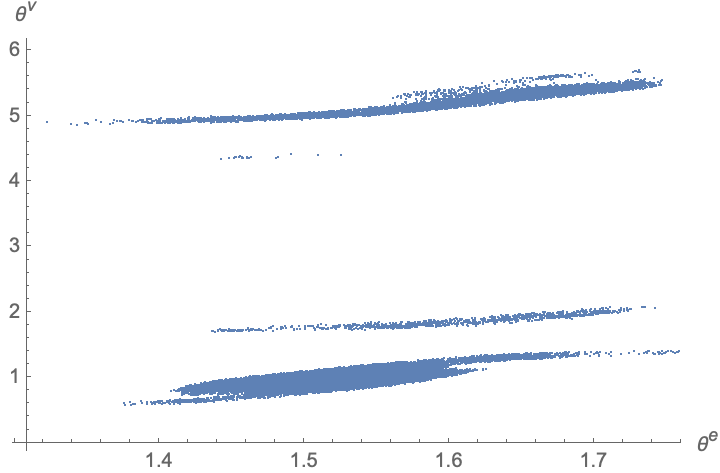}}  ~~~~
	\subfigure[flat pattern to PMNS in $\lambda_1$-$\lambda_2$]
		{\includegraphics[scale=0.25]{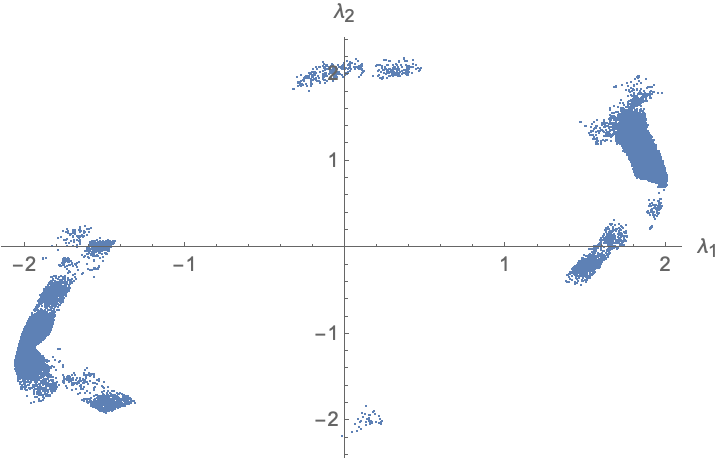}}
	\\
	\subfigure[seesaw pattern to CKM in $\theta^u$-$\theta^d$]
		{\includegraphics[scale=0.25]{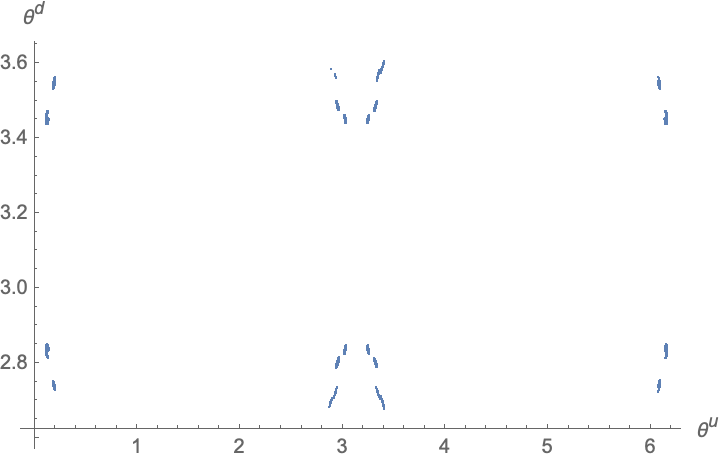}}  ~~~~
	\subfigure[seesaw pattern to CKM in $\lambda_1$-$\lambda_2$]
		{\includegraphics[scale=0.25]{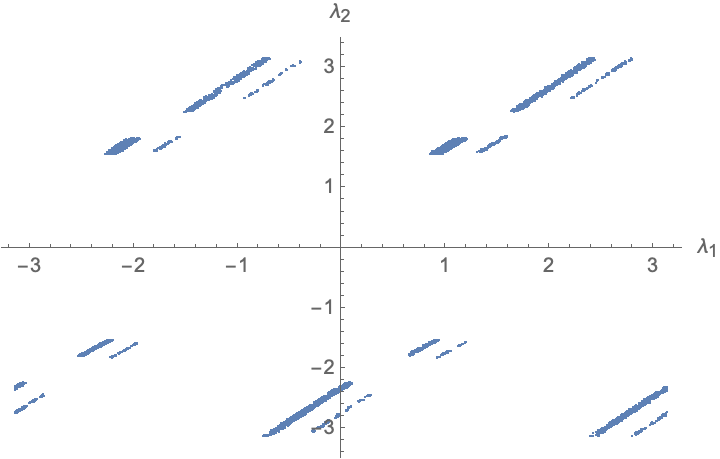}}
	\caption{ Global fit of flat pattern and seesaw pattern to CKM mixing and PMNS mixing. Mixing experiment data of quark and lepton are taken from \cite{PDG2022}. All fit points are in $2\sigma$ CL. The value of $m_1^\nu$ is input as $0.0001$ eV in subfigures (c) and (d).}
	\label{fig.fit}
\end{figure}

Quark/lepton mixings are determined by two rotation angles and two Yukawa phases. Scanning the space of four parameters, the allowed ranges are shown in Fig. \ref{fig.fit} for the flat pattern and seesaw pattern, respectively.
From Fig. \ref{fig.fit}, the parameter space for CKM mixing is markedly smaller than PMNS mixing due to more precise measurements of CKM mixing angles.

The flat pattern can completely pass tests of CKM mixing and PMNS mixing, while the seesaw pattern can not fit PMNS mixing. No parameter space is gotten for the seesaw pattern. Further calculations reveal that the seesaw pattern described in Equation (\ref{eq.ClostToseesawPMNSH1}) does not produce a sufficiently large value for $s^2_{23}$. Even with relaxed other PMNS mixing angles, the maximum of $s^2_{23}$ is only $0.0958$, far from the experiment value $0.547^{+0.018}_{-0.024}$.

In the above global fit, the lightest neutrino mass $m_1^\nu$ is initialed as $0.0001$ eV. The other two ones can be calculated from $\Delta m_{21}^2$ and $\Delta_{32}^2$ in the normal order. Because of the perturbative role of mass hierarchies, PMNS mixing angles are not sensitive to initialed value $m_1^\nu$, which has been verified by different initialing $m_1^\nu$. So, the failure of the seesaw pattern in lepton PMNS mixing can not be changed by initialed neutrino mass.

\section{Summary}
\label{sec.summary}

A common mass pattern for leptons and quarks has been decoded in terms of the hierarchy limit condition in the paper. 
As the phenomenological manifestation of flavor structure, the relation between flavor mixing and mass hierarchy has been understood from the common pattern for up-type and down-type quarks, charged leptons, and normal-order Dirac neutrino. 
The CKM mixing and PMNS mixing have the same mathematical structure and CP violations in the quark and lepton sectors all arise from Yukawa phases.
Mass hierarchy only plays a perturbative role in flavor mixing that is dominated by $SO(2)_L^f$ family symmetry and Yukawa phases.
Normal-order Dirac neutrinos become a necessary result of the common flavor structure for quarks and leptons.

Global fit results point out that a flat matrix is a successful common mass pattern of up-type, and down-type quarks, charged leptons, and neutrinos.
It hints at an equal and non-distinguished Yukawa interaction for the three families of quarks and leptons. 
In the hierarchy limit, the Yukawa term can be organized into a straightforward and unambiguous form
\begin{eqnarray*}
		-\mathcal{L}_Y
		=y^u\sum_{i,j}\bar{Q}_{L,i}^{(Y)}\tilde{H}\bar{u}_{R,j}^{(Y)}
			+y^d\sum_{i,j}\bar{Q}_{L,i}^{(Y)}{H}\bar{d}_{R,j}^{(Y)}
			+(u\rightarrow \nu, d\rightarrow e)
	\end{eqnarray*}
with a family universal coupling $y^f$ that determines the total family mass.

\begin{figure}[htbp]
	\centering  
	\includegraphics[scale=0.2]{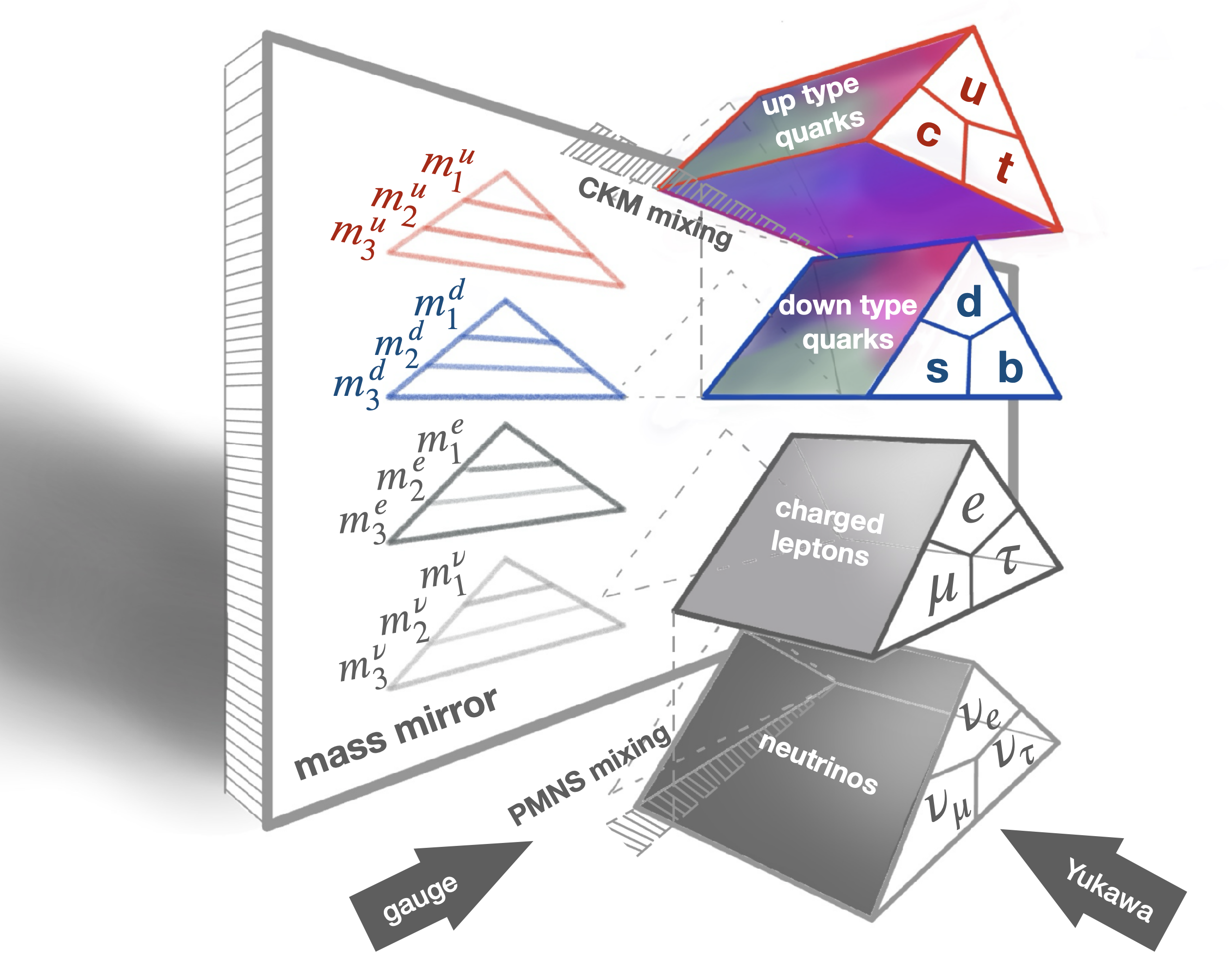}  
	\caption{Fundamental fermions}
	\label{fig.particle}
\end{figure}

Fundamental fermions can be re-arranged in terms of two view angles of gauge interaction and Yukawa interaction as shown in Fig. \ref{fig.particle}. 
The former is blind to the flavor structure, whereas the latter exhibits a non-distinguished interaction amongst three families on the Yukawa basis.
This figure explains the relationship between hierarchal masses and flat Yukawa interaction. Moreover, it reveals the cause of the quark/lepton mixing from the difference between two $SO(2)$ family rotations of two kinds of quarks/leptons.  
In the future, the remaining question to be answered regarding flavor is the dynamics of flavor breaking. It may arise from radiation correction of the SM or possible new physics beyond the SM.

\section*{Acknowledgements}
This work is partly supported by Shaanxi Natural Science Foundation 2022JM-052 and SAFS 22JSY035 of China.

\end{document}